\documentclass[
    ,final            % use final for the camera ready runs
  ]
  {aipproc}

\layoutstyle{6x9}

\usepackage{epsfig}
\usepackage{wrapfig,rotating}
\usepackage{amssymb}
\usepackage{amsmath}
\usepackage{hyperref}
\newcommand{\be}{\begin{equation}}
\newcommand{\ee}{\end{equation}}
\newcommand{\beq}{\begin{equation}}
\newcommand{\eeq}{\end{equation}}
\newcommand{\bea}{\begin{eqnarray}}
\newcommand{\eea}{\end{eqnarray}}

\newcommand{\mbf}[1]{\mathbf{#1}}

\begin{document}

\title{Dynamic versus Static Structure Functions \\ and Novel Diffractive Effects in QCD}

%\classification{ 12.38.Aw General properties of QCD,13.87.Fh Fragmentation into hadrons ,11.10.Kk
% Field theories in dimensions other than four,11.15.Tk Other nonperturbative techniques  }
%\

\author{Stanley J. Brodsky}{
  address={SLAC National Accelerator Laboratory
Stanford University, Stanford, CA 94309, USA}
 ,altaddress={Institute for Particle Physics and Phenomenology, Durham, UK}}

\begin{abstract}
Initial- and
final-state rescattering, neglected in the parton model, have a profound effect in QCD hard-scattering reactions,
predicting single-spin asymmetries, diffractive deep inelastic scattering, diffractive hard hadronic reactions, the breakdown of
the Lam Tung relation in Drell-Yan reactions, and nuclear shadowing and non-universal antishadowing---leading-twist physics not incorporated in
the light-front wavefunctions of the target computed in isolation. I also discuss the use of
diffraction to materialize the Fock states of a hadronic projectile and test QCD color transparency, and anomalous heavy quark effects. The
presence of direct higher-twist processes where a proton is produced in the hard subprocess can explain the large proton-to-pion ratio seen in
high centrality heavy ion collisions.  I emphasize the importance of distinguishing between static observables such as the probability distributions  computed from the square of the light-front wavefunctions versus dynamical observables which include the effects of rescattering. 
\end{abstract}

\maketitle

\section{Novel Features of Diffractive Deep Inelastic Scattering}

A remarkable feature of deep inelastic lepton-proton scattering at
HERA is that approximately 10\% events are
diffractive~\cite{Adloff:1997sc,Breitweg:1998gc}: the target proton
remains intact, and there is a large rapidity gap between the proton
and the other hadrons in the final state.  This observation presents a
paradox: if one chooses the conventional parton model frame, the virtual photon interacts with a quark constituent with
light-cone momentum fraction $x = {k^+/p^+} = x_{bj}.$  Furthermore,
the gauge link associated with the struck quark (the Wilson line)
becomes unity in light-cone gauge $A^+=0$. Thus the struck
``current'' quark apparently experiences no final-state
interactions. Since the light-front wavefunctions $\psi_n(x_i,
\mbf{k}_{\perp i})$ of a stable hadron are real, it appears
impossible to generate the required imaginary phase associated with
pomeron exchange, let alone large rapidity gaps. 
This paradox was resolved by Hoyer, Marchal,  Peigne, Sannino and myself~\cite{Brodsky:2002ue}.  Consider the case where the virtual photon
interacts with a strange quark---the $s \bar s$ pair is assumed to be produced in the target by gluon splitting.  In the case of Feynman gauge,
the struck $s$ quark continues to interact in the final state via gluon exchange as described by the Wilson line. The final-state interactions
occur at a light-cone time $\Delta\tau \simeq 1/\nu$ shortly after the virtual photon interacts with the struck quark. When one integrates over
the nearly-on-shell intermediate state, the amplitude acquires an imaginary part. Thus the rescattering of the quark produces a separated
color-singlet $s \bar s$ and an imaginary phase. In the case of the light-cone gauge $A^+ = \eta \cdot A =0$, one must also consider the
final-state interactions (rescattering)  of the (unstruck) $\bar s$ quark. The gluon propagator in light-cone gauge is inversely proportional to  $k^+ .$ The
momentum of the exchanged gluon $k^+$ is of ${ \cal O}{(1/\nu)}$; thus rescattering contributes at leading twist even in light-cone gauge. The
net result is  gauge invariant and identical to a color dipole model calculation. The calculation of the rescattering effects on DIS in
Feynman and light-cone gauge through three loops is given in detail for an Abelian model in ref.~\cite{Brodsky:2002ue}.  The result shows
that the rescattering corrections reduce the magnitude of the DIS cross section in analogy to nuclear shadowing.

A new understanding of the role of rescattering in deep inelastic
scattering has thus emerged. The multiple scattering of the struck
parton via instantaneous interactions in the target generates
dominantly imaginary diffractive amplitudes, giving rise to an
effective ``hard pomeron'' exchange.  The presence of a rapidity gap
between the target and diffractive system requires that the target
remnant emerges in a color-singlet state; this is made possible in
any gauge by the soft rescattering.  The resulting diffractive
contributions leave the target intact  and do not resolve its quark
structure; thus there are contributions to the DIS structure
functions which cannot be interpreted as parton
probabilities~\cite{Brodsky:2002ue}; the leading-twist contribution
to DIS  from rescattering of a quark in the target is a coherent
effect which is not included in the light-front wavefunctions
computed in isolation. 

Another novel  QCD phenomenon involving nuclei is the {\it antishadowing} of the nuclear structure functions which is observed in deep
inelastic lepton scattering and other hard processes. Empirically, one finds $R_A(x,Q^2) \equiv  \left(F_{2A}(x,Q^2)/ (A/2) F_{d}(x,Q^2)\right)
> 1 $ in the domain $0.1 < x < 0.2$; {\em i.e.}, the measured nuclear structure function (referenced to the deuteron) is larger than than the
scattering on a set of $A$ independent nucleons.
Ivan Schmidt, Jian-Jun Yang, and I~\cite{Brodsky:2004qa} have extended the analysis of nuclear shadowing  to the shadowing and antishadowing of all of the
electroweak structure functions.  We note that there are also leading-twist diffractive contributions $\gamma^* N_1 \to (q \bar q) N_1$  arising from Reggeon exchanges in the
$t$-channel~\cite{Brodsky:1989qz}.  For example, isospin--non-singlet $C=+$ Reggeons contribute to the difference of proton and neutron
structure functions, giving the characteristic Kuti-Weisskopf $F_{2p} - F_{2n} \sim x^{1-\alpha_R(0)} \sim x^{0.5}$ behavior at small $x$. The
$x$ dependence of the structure functions reflects the Regge behavior $\nu^{\alpha_R(0)} $ of the virtual Compton amplitude at fixed $Q^2$ and
$t=0.$ The phase of the diffractive amplitude is determined by analyticity and crossing to be proportional to $-1+ i$ for $\alpha_R=0.5,$ which
together with the phase from the Glauber cut, leads to {\it constructive} interference of the diffractive and nondiffractive multi-step nuclear
amplitudes.  The nuclear structure function is predicted to be enhanced precisely in the domain $0.1 < x <0.2$ where
antishadowing is empirically observed.  The strength of the Reggeon amplitudes is fixed by the fits to the nucleon structure functions, so there
is little model dependence.
Quarks of different flavors  will couple to different Reggeons; this leads to the remarkable prediction that
nuclear antishadowing is not universal; it depends on the quantum numbers of the struck quark. This picture implies substantially different
antishadowing for charged and neutral current reactions, thus affecting the extraction of the weak-mixing angle $\theta_W$.  We find that part
of the anomalous NuTeV result~\cite{Zeller:2001hh} for $\theta_W$ could be due to the non-universality of nuclear antishadowing for charged and
neutral currents.  In fact,  Schienbein et al.~\cite{Schienbein:2008ay} have recently given a comprehensive analysis of charged current deep inelastic neutrino-iron scattering, finding significant differences with the nuclear corrections for electron-iron scattering.

\begin{figure}[!]
 %\begin{center}
\includegraphics[width=13cm]{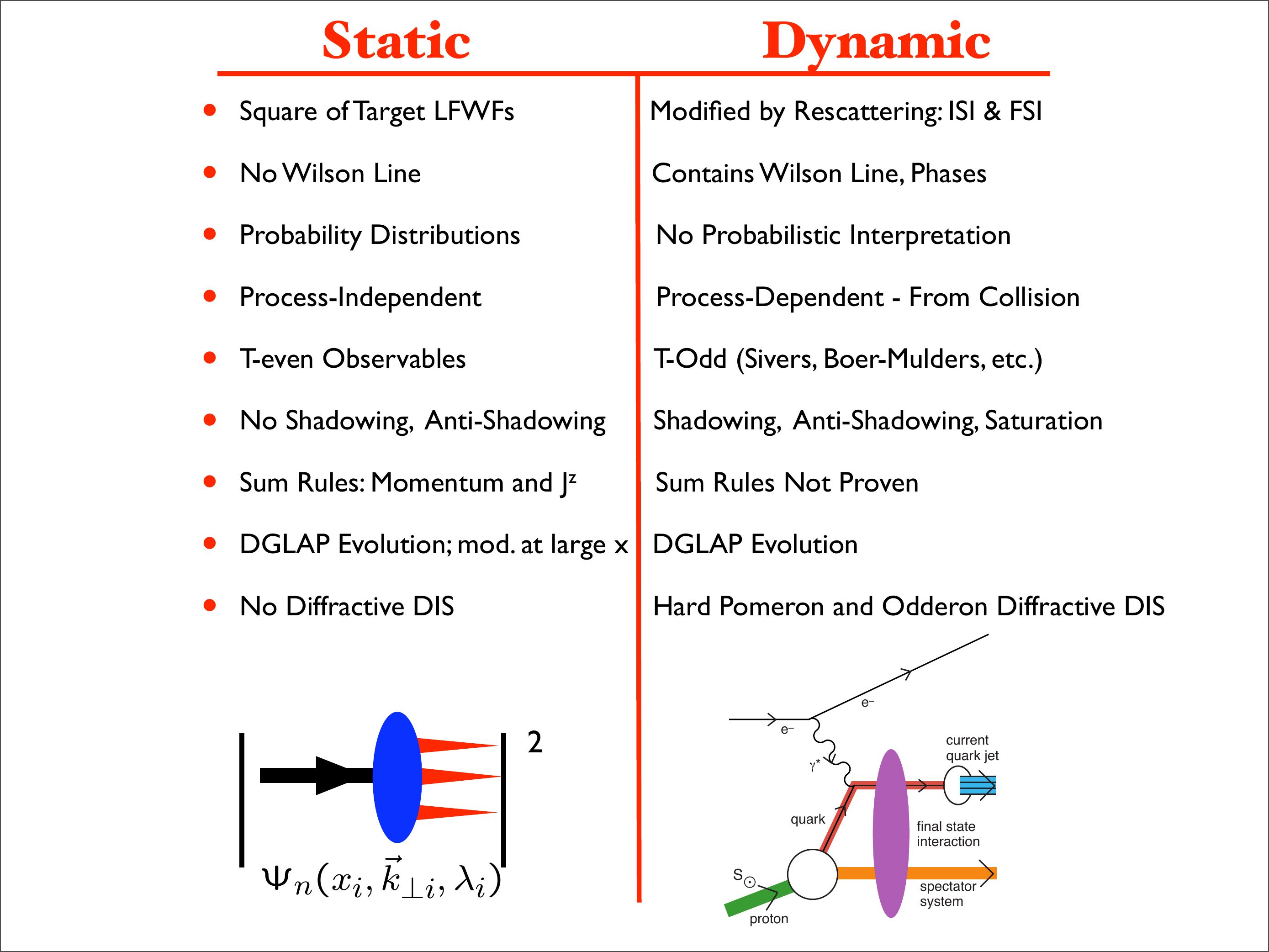}
%\end{center}
%\vspace{-30pt}
\caption{Static versus dynamic structure functions}
\label{figstatdyn}  
\end{figure}

It is thus important to distinguish ``static" structure functions which are computed directly from the LFWFs of the target  from the ``dynamic" empirical structure functions which take into account rescattering of the struck quark.   Since they derive from the LF eigenfunctions of the target hadron, the static structure functions have a probabilistic interpretation.  Since the wavefunction of a stable eigenstate is real, the static structure functions do not describe DDIS nor the single-spin asymmetries discussed below since such phenomena involves the complex phase structure of the $\gamma^* p $ amplitude.  
One can augment the light-front wavefunctions with a gauge link corresponding to an external field
created by the virtual photon $q \bar q$ pair
current~\cite{Belitsky:2002sm,Collins:2004nx}, but such a gauge link is
process dependent~\cite{Collins:2002kn}, so the resulting augmented
wavefunctions are not universal.
\cite{Brodsky:2002ue,Belitsky:2002sm,Collins:2003fm}. We  emphasize
that the shadowing of nuclear structure functions is due to the
destructive interference between multi-nucleon amplitudes involving
diffractive DIS and on-shell intermediate states with a complex
phase.  The physics of rescattering and shadowing is thus not
included in the nuclear light-front wavefunctions, and a
probabilistic interpretation of the nuclear DIS cross section is
precluded. 
The distinction 
between static structure functions; i.e., the probability distributions  computed from the square of the light-front wavefunctions, versus the nonuniversal dynamic structure functions measured in deep inelastic scattering is summarized in fig. \ref{figstatdyn}.

\section{ Single-Spin Asymmetries  and Other Leading-Twist  Rescattering Effects}

Among the most interesting polarization effects are single-spin
azimuthal asymmetries  in semi-inclusive deep inelastic scattering,
representing the correlation of the spin of the proton target and
the virtual photon to hadron production plane: $\vec S_p \cdot \vec
q \times \vec p_H$. Such asymmetries are time-reversal odd, but they
can arise in QCD through phase differences in different spin
amplitudes. In fact, final-state interactions from gluon exchange
between the outgoing quarks and the target spectator system lead to
single-spin asymmetries (SSAs) in semi-inclusive deep inelastic
lepton-proton scattering  which  are not power-law suppressed at
large photon virtuality $Q^2$ at fixed
$x_{bj}$~\cite{Brodsky:2002cx}.  In contrast to the SSAs arising
from transversity and the Collins fragmentation function, the
fragmentation of the quark into hadrons is not necessary; one
predicts a correlation with the production plane of the quark jet
itself. Physically, the final-state interaction phase arises as the
infrared-finite difference of QCD Coulomb phases for hadron 
wavefunctions with differing orbital angular momentum.  The same proton
matrix element which determines the spin-orbit correlation $\vec S
\cdot \vec L$ also produces the anomalous magnetic moment of the
proton, the Pauli form factor, and the generalized parton
distribution $E$ which is measured in deeply virtual Compton
scattering. Thus the contribution of each quark current to the SSA
is proportional to the contribution $\kappa_{q/p}$ of that quark to
the proton target's anomalous magnetic moment $\kappa_p = \sum_q e_q
\kappa_{q/p}$ ~\cite{Brodsky:2002cx,Burkardt:2004vm}.  The HERMES
collaboration has recently measured the SSA in pion
electroproduction using transverse target
polarization~\cite{Airapetian:2004tw}.  A related analysis also
predicts that the initial-state interactions from gluon exchange
between the incoming quark and the target spectator system lead to
leading-twist single-spin asymmetries in the Drell-Yan process $H_1
H_2^\updownarrow \to \ell^+ \ell^- X$
~\cite{Collins:2002kn,Brodsky:2002rv}.  The SSA in the Drell-Yan
process is the same as that obtained in SIDIS, with the appropriate
identification of variables, but with the opposite sign. There is no such
single spin asymmetries in charged-current reactions since the $W$ only
couples to left-handed quarks~\cite{Brodsky:2002pr}. If both the
quark and antiquark in the initial state of the Drell-Yan subprocess
$q \bar q \to  \mu^+ \mu^-$ interact with the spectators of the
other incident hadron, one finds a breakdown of the Lam-Tung
relation, which was formerly believed to be a general prediction of
leading-twist QCD. These double initial-state interactions also lead
to a $\cos 2 \phi$ planar correlation in unpolarized Drell-Yan
reactions~\cite{Boer:2002ju}.

As noted by Collins and Qiu~\cite{Collins:2007nk}, the traditional factorization formalism of perturbative QCD for high transverse
momentum hadron production  fails in detail even at the LHC because of initial- and final-state rescattering.  The signal for factorization breakdown is a $\cos 2 \phi$ planar correlation in dijet production.

\section{Novel Heavy Quark Phenomena}

The probability for Fock states of a light hadron such as the proton to have an extra heavy quark pair decreases as $1/m^2_Q$ in non-Abelian
gauge theory~\cite{Franz:2000ee,Brodsky:1984nx}.  The relevant matrix element is the cube of the QCD field strength $G^3_{\mu \nu}.$  This is in
contrast to abelian gauge theory where the relevant operator is $F^4_{\mu \nu}$ and the probability of intrinsic heavy leptons in QED bound
state is suppressed as $1/m^4_\ell.$  The intrinsic Fock state probability is maximized at minimal off-shellness.  It is useful to define the
transverse mass $m_{\perp i}= \sqrt{k^2_{\perp i} + m^2_i}.$ The maximum probability then occurs at $x_i = { m^i_\perp /\sum^n_{j = 1}
m^j_\perp}$; {\em i.e.}, when the constituents have minimal invariant mass and equal rapidity. Thus the heaviest constituents have the highest
momentum fractions and the highest $x_i$. Intrinsic charm thus predicts that the charm structure function has support at large $x_{bj}$ in
excess of DGLAP extrapolations~\cite{Brodsky:1980pb}; this is in agreement with the EMC measurements~\cite{Harris:1995jx}. Intrinsic charm can
also explain the $J/\psi \to \rho \pi$ puzzle~\cite{Brodsky:1997fj}. It also affects the extraction of suppressed CKM matrix elements in $B$
decays~\cite{Brodsky:2001yt}.
The dissociation of the intrinsic charm $|uud c \bar c>$ Fock state of the proton on a nucleus can produce a leading heavy quarkonium state at
high $x_F = x_c + x_{\bar c}~$ in $p A \to J/\psi X A^\prime$ since the $c$ and $\bar c$ can readily coalesce into the charmonium state.  Since
the constituents of a given intrinsic heavy-quark Fock state tend to have the same rapidity, coalescence of multiple partons from the projectile
Fock state into charmed hadrons and mesons is also favored.  For example, one can produce a leading $\Lambda_c$ at high $x_F$ and low $p_T$ from
the coalescence of the $u d c$ constituents of the projectile $|uud c \bar c>$  Fock state. 

In the case of a nuclear target, the charmonium state will be produced at small transverse momentum and high $x_F$  with a characteristic
$A^{2/3}$ nuclear dependence since the color-octet color-octet $|(uud)_{8C} (c \bar c)_{8C} >$ Fock state interacts on the front surface of the
nuclear target~\cite{Brodsky:2006wb}. This forward contribution is in addition to the $A^1$ contribution derived from the usual perturbative QCD
fusion contribution at small $x_F.$   Because of these two components, the cross section violates perturbative QCD factorization for hard
inclusive reactions~\cite{Hoyer:1990us}.  This is consistent with the observed two-component cross section for charmonium production observed by
the NA3 collaboration at CERN~\cite{Badier:1981ci} and more recent experiments~\cite{Leitch:1999ea}. The diffractive dissociation of the
intrinsic charm Fock state leads to leading charm hadron production and fast charmonium production in agreement with
measurements~\cite{Anjos:2001jr}.  The hadroproduction cross sections for  double-charm $\Xi_{cc}^+$ baryons at SELEX~\cite{Ocherashvili:2004hi} and the production of $J/\psi$ pairs at NA3 are
be consistent with the diffractive dissociation and coalescence of double IC Fock states~\cite{Vogt:1995tf}. These observations provide
compelling evidence for the diffractive dissociation of complex off-shell Fock states of the projectile and contradict the traditional view that
sea quarks and gluons are always produced perturbatively via DGLAP evolution. It is also conceivable that the observations~\cite{Bari:1991ty} of
$\Lambda_b$ at high $x_F$ at the ISR in high energy $p p$  collisions could be due to the diffractive dissociation and coalescence of the
``intrinsic bottom" $|uud b \bar b>$ Fock states of the proton.
As emphasized by Lai, Tung, and Pumplin~\cite{Pumplin:2007wg}, there are strong indications that the structure functions used to model charm
and bottom quarks in the proton at large $x_{bj}$ have been strongly underestimated, since they ignore intrinsic heavy quark fluctuations of
hadron wavefunctions.   

Goldhaber, Kopeliovich, Schmidt, Soffer, and I ~\cite{Brodsky:2006wb,Brodsky:2007yz} have  proposed a novel  mechanism for exclusive diffractive
Higgs production $pp \to p H p $  and nondiffractive Higgs production  in which the Higgs boson carries a significant fraction of the projectile proton momentum. The production
mechanism is based on the subprocess $(Q \bar Q) g \to H $ where the $Q \bar Q$ in the $|uud Q \bar Q>$ intrinsic heavy quark Fock state has up
to $80\%$ of the projectile protons momentum. This mechanism provides a clear experimental signal 
for Higgs production at the LHC due to the small background in this kinematic region.

\section{Diffraction Dissociation as a Tool to Resolve Hadron
Substructure and Test Color Transparency}

Diffractive multi-jet production in heavy nuclei provides a novel way to resolve the shape of light-front Fock state wavefunctions and test
color transparency~\cite{Brodsky:1988xz}.  For example, consider the reaction~\cite{Bertsch:1981py,Frankfurt:1999tq}.   $\pi A \rightarrow {\rm
Jet}_1 + {\rm Jet}_2 + A^\prime$ at high energy where the nucleus $A^\prime$ is left intact in its ground state. The transverse momenta of the
jets balance so that $ \vec k_{\perp i} + \vec k_{\perp 2} = \vec q_\perp < {R^{-1}}_A  .$  Because of color transparency, the valence wavefunction of the pion with small impact separation will penetrate the nucleus with minimal interactions, diffracting into jet
pairs~\cite{Bertsch:1981py}.  The $x_1=x$, $x_2=1-x$ dependence of the dijet distributions will thus reflect the shape of the pion valence
light-cone wavefunction in $x$; similarly, the $\vec k_{\perp 1}- \vec k_{\perp 2}$ relative transverse momenta of the jets gives key
information on the second transverse momentum derivative of the underlying shape of the valence pion
wavefunction~\cite{Frankfurt:1999tq,Nikolaev:2000sh}. The diffractive nuclear amplitude extrapolated to $t = 0$ should be linear in nuclear
number $A$ if color transparency is correct.  The integrated diffractive rate will then scale as $A^2/R^2_A \sim A^{4/3}$. This is in fact what
has been observed by the E791 collaboration at FermiLab for 500 GeV incident pions on nuclear targets~\cite{Aitala:2000hc}. 

Light-Front Holography is one of the most remarkable features of AdS/CFT~\cite{deTeramond:2008ht,Brodsky:2008gc,Brodsky:2008td}.  It  allows one to project the functional dependence of the wavefunction $\Phi(z)$ computed  in the  AdS fifth dimension to the  hadronic frame-independent light-front wavefunction $\psi(x_i, \mbf{b}_{\perp i})$ in $3+1$ physical space-time. The
variable $z $ maps  to $ \zeta(x_i, \mbf{b}_{\perp i})$. To prove this, we have shown that there exists a correspondence between the matrix elements of the energy-momentum tensor of the fundamental hadronic constituents in QCD with the transition amplitudes describing the interaction of string modes in anti-de Sitter space with an external graviton field which propagates in the AdS interior. The agreement of the results for both electromagnetic and gravitational hadronic transition amplitudes provides an important consistency test and verification of holographic mapping from AdS to physical observables defined on the light-front~\cite{Brodsky:2008pf}.   In fact $\zeta$ is the only variable to appear in the light-front
Schr\"odinger equations predicted from AdS/QCD~\cite{deTeramond:2008ht}.  These equations for both meson and baryons give a good representation of the observed hadronic spectrum, especially in the case of the soft wall model. The resulting LFWFs also have excellent phenomenological features, including predictions for the  electromagnetic form factors and decay constants.  We have also shown that the LF Hamiltonian formulation of quantum field theory provides a natural formalism to compute
hadronization at the amplitude level~\cite{Brodsky:2008tk}.
It is interesting to note that the form of the nonperturbative pion distribution amplitude $ \phi_\pi(x)$ obtained from integrating the prediction from AdS/QCD $ q \bar q$ valence LFWF $\psi(x, \mbf{k}_\perp)$  over $\mbf{k}_\perp$,
has a quite different $x$-behavior~\cite{Brodsky:2006uqa} than the
asymptotic distribution amplitude predicted from  PQCD
evolution~\cite{Lepage:1979zb}.
The AdS/QCD prediction
$ \phi_\pi(x)  = \sqrt{3}  f_\pi \sqrt{x(1-x)}$ has a broader distribution than expected from solving the ERBL evolution equation in perturbative QCD.
This observation appears to be consistent with the results of the Fermilab diffractive dijet
experiment~\cite{Aitala:2000hb} in the low $p_T$ regime, the moments obtained from lattice QCD~\cite{Brodsky:2008pg} and pion form factor data~\cite{Choi:2006ha}.

\section{Color Transparency and the RHIC Baryon Anomaly}

It is conventional to assume that leading-twist subprocesses dominate measurements of high $p_T$ hadron production at RHIC energies. Indeed the
data for direct photon fragmentation $ p p \to \gamma X $ is quite consistent with $n_{eff}(p p \to \gamma X) = 5 ,$  as expected from the $g q
\to \gamma q$ leading-twist subprocess. This also is likely true for pion production, at least for small $x_T.$  However, the measured fixed
$x_T$ scaling for proton production at RHIC is anomalous:  PHENIX reports $n_{eff} (p p \to p X)\simeq 8$.  A review of this data is given by
Tannenbaum~\cite{Tannenbaum:2006ku}. One can understand the anomalous scaling if a higher-twist subprocess~\cite{Brodsky:2008qp} where the
proton is made {\it directly} within the hard reaction, such as $ u u \to p \bar d$ and $(uud) u \to p u$, dominates the reaction $ p p \to p X$
at RHIC energies.  
Such processes are rigorous QCD contributions. The dominance of direct subprocesses is possible since the fragmentation of gluon or quark jets
to baryons requires that the  2 to 2 subprocess occurs at much higher transverse momentum than the $p_T$ of observed proton because of the fast-falling $(1-z)^3 $ quark-to-proton fragmentation function.  Such ``direct" reactions can readily explain the fast-falling power-law  falloff
observed at fixed $x_T$ and fixed-$\theta_{cm}$ observed at the ISR, FermiLab and RHIC.
Furthermore, the protons produced
directly within the hard subprocess emerge as small-size color-transparent colored states which are not absorbed in the nuclear target. In
contrast, pions produced from jet fragmentation have the normal cross section. This provides a plausible explanation of the RHIC
data,~\cite{Adler:2003kg} which shows a dramatic rise of the $p \to \pi$ ratio at high $p_T$ when one compares  peripheral  with  central (full
overlap) heavy ion collisions.  The directly produced protons are not absorbed, but the pions are
diminished in the nuclear medium.

\section*{Acknowledgments}
Presented  at 
Diffraction 2008: International Workshop On Diffraction In High Energy Physics
9-14 Sep 2008, La Londe-les-Maures, France.  I thank  Paul Hoyer, Ivan Schmidt, Boris Kopeliovich, Dae Sung Hwang, Fred Goldhaber, Jacques Soffer, Anne Sickles, Ramona Vogt, Daniel Boer and Guy de Teramond for our collaborations.
I also am grateful to  Jacques Soffer, and Roberto Fiore for their hospitality at Diffraction 2008.   SLAC-PUB-13454
This research was supported by the Department
of Energy contract DE--AC02--76SF00515.

\end{document}